# Evaluating the efficacy of haptic feedback, 360° treadmill-integrated Virtual Reality framework and longitudinal training on decision-making performance in a complex search-and-shoot simulation

Akash K Rao*[a], Arnav Bhavsar[b], Shubhajit Roy Chowdhury[b], Sushil Chandra[c], Ramsingh Negi[c] Prakash Duraisamy[d], and Varun Dutt[a]

[a]Applied Cognitive Science Laboratory, Indian Institute of Technology Mandi, Himachal Pradesh, India; [b]School of Computing and Electrical Engineering, Indian Institute of Technology Mandi, Himachal Pradesh, India; [c]Institute of Nuclear Medicine and Allied Sciences, Defence Research and Development Organization, Delhi, India; [d]University of Wisconsin-Green Bay, USA

**ABSTRACT**

Virtual Reality (VR) has made significant strides, offering users a multitude of ways to interact with virtual environments. Each sensory modality in VR provides distinct inputs and interactions, enhancing the user's immersion and presence. However, the potential of additional sensory modalities, such as haptic feedback and 360° locomotion, to improve decision-making performance has not been thoroughly investigated. This study addresses this gap by evaluating the impact of a haptic feedback, 360° locomotion-integrated VR framework and longitudinal, heterogeneous training on decision-making performance in a complex search-and-shoot simulation. The study involved 32 participants from a defence simulation base in India, who were randomly divided into two groups: experimental (haptic feedback, 360° locomotion-integrated VR framework with longitudinal, heterogeneous training) and placebo control (longitudinal, heterogeneous VR training without extrasensory modalities). The experiment lasted 10 days. On Day 1, all subjects executed a search-and-shoot simulation closely replicating the elements/situations in the real world. From Day 2 to Day 9, the subjects underwent heterogeneous training, imparted by the design of various complexity levels in the simulation using changes in behavioral attributes/artificial intelligence of the enemies. On Day 10, they repeated the search-and-shoot simulation executed on Day 1. The results showed that the experimental group experienced a gradual increase in presence, immersion, and engagement compared to the placebo control group. However, there was no significant difference in decision-making performance between the two groups on day 10. We intend to use these findings to design multisensory VR training frameworks that enhance engagement levels and decision-making performance.

**Keywords:** Virtual Reality, Haptic Feedback, 360° Locomotion Treadmill, Heterogenous Training, Presence

## 1. INTRODUCTION

Serious simulations, like first-person search-and-shoot games, place a strong emphasis on perceptual processing, attentional control, spatial cognition, and rapid decision-making [1]. These simulations demand that individuals evaluate multiple stimuli, prioritize targets, formulate strategies, and perform precise actions within tight deadlines and varying levels of complexity [1]. The cognitive demands of these simulations activate neural pathways linked with visual-spatial processing, executive functions, and response inhibition, which fosters the development of adaptive cognitive strategies and improved performance over time [2,3]. The capacity of the brain to adapt and reorganize in response to experiences and learning, known as neuroplasticity, is a crucial mechanism through which serious simulations enhance decision-making [2]. Engaging regularly in action-packed serious simulations prompts neuroplastic changes, especially in regions associated with attention, working memory, and decision-making [4]. These changes result in better cognitive skills, such as quicker reaction times, heightened situational awareness, and more accurate decision-making [4]. Studies in neurocognition have shown that training in serious simulations leads to alterations in the brain's structure and function, including increased grey matter density in prefrontal and parietal regions responsible for executive control and attentional processes [5]. Previous research indicates that skills honed through gameplay, such as attentional focus, multitasking, pattern recognition, and decision-making under pressure, can be applied to non-gaming contexts [6]. This transferability is attributed to the cognitive processes and strategies developed during gameplay, which can be utilized in real-life situations that require similar cognitive demands [6].

Unlike conventional serious simulations that rely solely on non-immersive visual and auditory stimuli, Virtual Reality (VR) has the potential to incorporate a variety of sensory modalities to create a multisensory environment. This heightened sense of presence, characterized by a feeling of "being there" in the virtual world, has been shown to have critical implications for cognitive processing and decision-making [5,7]. Research studies on VR-based serious simulations have emphasized the unique advantages of immersive environments on cognitive engagement [5,6-8]. By placing users directly within the game world, VR has been demonstrated to enhance spatial cognition, situational awareness, and emotional involvement [8]. These factors are essential for decision-making, allowing individuals to assess complex scenarios from multiple perspectives, anticipate outcomes, and adjust strategies accordingly [9]. Additionally, VR enables embodied cognition, where users interact with virtual objects and environments through naturalistic movements [10,11]. This embodied interaction reinforces learning through physical actions and strengthens the sense of agency and control critical for effective decision-making. Consequently, VR-based serious simulations have served as immersive training grounds that challenge cognitive processes and nurture adaptive decision-making skills [12].

The advancement of virtual reality (VR) experiences depends on incorporating additional sensory modalities, such as haptic feedback systems, critical for enhancing immersion, realism, and user experience within VR environments [13]. Haptic feedback provides a crucial dimension of realism by simulating physical interactions with virtual objects and surfaces, unlike traditional visual and auditory cues. Research has demonstrated that haptic cues can influence decision-making by providing additional contextual information and reinforcing intuitive judgments, particularly in dynamic and complex scenarios [14-16]. In action-oriented simulations, such as the search-and-shoot simulation investigated in this study, haptic feedback can alert users to environmental cues, simulate weapon recoil, or convey spatial constraints, all impacting decision-making strategies [15,16]. Research studies exploring the impact of haptic feedback on VR experiences have highlighted its profound effects on immersion and presence [17-19]. By providing users with tactile feedback corresponding to virtual interaction, haptic feedback systems have been reported to create a significantly enhanced and cohesive sensory experience, aligning visual, auditory, and tactile stimuli to create a sense of "being there" in the virtual environment, eventually contributing to a heightened sense of realism, emotional engagement, and cognitive processing within VR scenarios [18,20]. Researchers in [21] conveyed that haptic feedback can impart knowledge regarding sensations of weight, texture, and resistance, thereby improving motor skills and spatial awareness. However, some researchers have also highlighted the limitations of modern haptic feedback systems, including the need for higher fidelity haptic devices, improved synchronization with visual and auditory stimuli, and considerations for user comfort and ergonomics [22-24]. Despite the research on the ergonomic and usability issues associated with modern haptic feedback systems, their concentrated effect on presence and enhancement in decision-making performance, especially when used as an effective supplement to typical visuo-auditory VR simulations, requires further investigation. However, research on the behavioral implications of using haptic feedback systems compared to conventional visuo-auditory VR modalities and their effects on presence, simulator sickness, and performance is currently lacking in the literature.

Like haptic feedback systems, 360° locomotion treadmills signify a notable progression in virtual reality interaction, facilitating users to engage in naturalistic movement within virtual spaces [25]. In contrast to traditional VR setups that rely on handheld controllers for locomotion, these treadmills allow users to walk, run, turn, and navigate freely in all directions, enabling unrestricted movement [25-27]. The ability to move freely and interact with virtual spaces as they would in the physical world eliminates the disconnect between motor actions and visual feedback, reducing motion sickness and enhancing spatial presence [27]. This heightened sense of immersion fosters a stronger emotional connection to the virtual environment, promoting engagement and investment in decision-making tasks [28]. Previous research has suggested that physical immersion and naturalistic movement in VR can lead to more intuitive and contextually grounded decision-making [29]. In tasks requiring spatial navigation, strategic planning, and quick responses, such as the search-and-shoot simulation mentioned in this research work, 360° treadmills enable users to adopt real-world strategies and movements, enhancing decision-making efficiency [30]. Researchers in [31] have also investigated that the proprioceptive feedback provided by 360º locomotion treadmills contributes to cognitive load management, where motor actions and spatial awareness offload cognitive processing and support more streamlined decision-making. However, similar to the haptic feedback systems, research on how locomotion treadmills act as an effective addition to increasing immersion and performance in dynamic decision-making tasks is lacking and much needed in the literature.

Beyond the immediate impact of immersive technologies, longitudinal training plays a crucial role in sustaining and reinforcing cognitive gains over time [32]. Previous research on longitudinal training in VR has demonstrated promising results in various domains, from surgical skills acquisition to military training scenarios [33,34]. In addition, researchers have also evaluated the efficiency of different longitudinal training conditions on spatial navigation and motor coordination tasks [34]. These longitudinal training frameworks include the 'retrieval effort hypothesis' [35], 'cognitive antidote hypothesis' [35], 'procedural reinstatement' [35], 'heterogeneity by practice' [35], among others. Among these training frameworks, the 'heterogeneity by practice' training framework has been found to be the most effective due to its propensity to store effective, variable instances of the mental model required to successfully circumvent the objectives in a dynamic decision-making task [35]. However, the efficacy of this framework has yet to be tested in dynamic decision-making simulations, especially when extrasensory modalities have been inserted into the conventional visual-auditory VR testbeds for increased immersion. In this research work, we try to address this gap in the literature by designing an experiment to evaluate the impact of haptic feedback, 360° locomotion-integrated VR framework and longitudinal, heterogeneous training on decision-making performance in a complex search-and-shoot simulation. In what follows, we first describe an experiment to investigate the efficacy of haptic feedback, 360° locomotion-integrated VR framework on decision-making performance in a serious simulation. We then discuss the implications of the results obtained from the perspective of designing VR frameworks embellished with extrasensory modalities for enhancing immersion, presence, and decision-making performance.

## 2. MATERIALS AND METHODS

### 2.1 Participants

Thirty-two participants (All males, 30 right-handed, mean age = 26.7 years, SD = 2.3 years) at a defence research simulation base in India participated in the study. The ethical committees at the Indian Institute of Technology Mandi and the defence research simulation base approved the experiment. All the subjects were from a military background, and the recruitment for the study was done through the pool of personnel available at the defence research simulation base in compliance with the inclusion criteria. None of the participants reported any history of psychological/mental disorders, and all participants had either normal or corrected to normal vision. Inclusion criteria included questionnaires concerning their demographic details, a sleepiness scale (recorded through the Epworth Sleepiness Scale [36]), pre-simulator sickness (details regarding history of nausea, headaches, and recurring migraines), and a visual acuity test. All the participants reported that they had little to no experience with VR, and they had no experience with haptic feedback systems and 360° locomotion systems. All the participants received a flat payment of INR 4000 for their participation in the study, and the top-performing participants (3 of the 32 who participated in the experiment) received a bonus of INR 5000 in addition to the participation fee.

### 2.2 Search-and-shoot simulation

The search-and-shoot simulation was designed using Unreal Engine 5, with the virtual characters and the gun controller (M4 carbine 5.56mm) being designed using Blender Animation v 2.8. As shown in Figure 1(a), the search-and-shoot simulation consisted of a typical maze-like urban landscape consisting of four interconnected streets, alleys, and buildings, reflecting elements of urban warfare. The environment comprised dilapidated buildings, alleyways strewn with debris, and a few vantage points such as rooftops and balconies. The color palette included shades of brown and grey to reflect the ambience of a conflict zone. Before the experiment commenced, the participants were instructed that the enemies in the serious simulation had invaded three different security bases in the physical mesh. The participant's objective was to strategically kill all enemies and reacquire all the security bases in the physical mesh within the stipulated time of 5 minutes. The participant's health in the simulation was initialized at 100. The subsequent decrease in health in the simulation was based on the complexity levels programmed in the simulation (refer to section 2.4). The total number of enemies in the simulation was initialized to 9, where any number of enemies would be randomly allocated at a given security base. Five different levels of simulation complexity (rookie, intermediate–level 1 (L1), intermediate–level 2 (L2), professional, and transfer) were designed. The 'transfer' difficulty level was only used on Days 1 and 10 to test the efficacy of the training interventions, and all the other levels of complexity were used during the training intervention. The complexity levels were incorporated by tweaking some basic physical characteristics of the enemy avatars and the participants (refer to section 2.4). As shown in Figure 1(b), the VR-based search-and-shoot serious simulation was executed using a HTC Vive Pro, rendering the simulation at a resolution of 1440 x 1600 pixels per eye at a refresh rate of around 90 Hz. The SteamVR tracking sensors were incorporated into the simulation using the OpenXR framework for movement and shooting. An NVIDIA GeForce RTX 3090Ti GPU, with 256 GB RAM and an AMD Ryzen Threadripper 3990X CPU, was used to render the simulation seamlessly. For haptic (vibrotactile) feedback, we used the Tactsuit X40 from bHaptics [37]. This haptic suit consisted of 40 haptic feedback (Eccentric rotating mass vibrotactile motors), with 20 motors on the front side of the vest and 20 on the backside. The haptic vest consisted of highly flexible shoulder snap buttons with a lot of room for adjustment with respect to the participant's size and chest. In addition, the participants also wore arm sleeves between the wrist and the elbow with adjustable straps. These arm sleeves consisted of 6 ERM vibration motors each. We used the bHaptics plugin to incorporate the Tactsuit X40 into the search-and-shoot simulation. The haptic events (which included gunshots from the enemies, recoil feedback from the gun, among others) were deployed using the bHaptics developer framework in Unreal Engine 5. As seen in Figure 1(a), we used the Kat Walk C (size-large) locomotive platform to integrate physical movement in the simulation. The Kat Walk C was seamlessly integrated into the simulation using the KatVR Unreal integration framework. The KatVR sensors are connected to this integration framework to receive the participant's operations in real-time. The locomotion tracking was executed by three sensors in the locomotion platform, with one at the torso of the participant (to track the direction of the movement) and two on the participant's shoes (to track the actual movement). The experiment was conducted in an isolated location with limited noise.

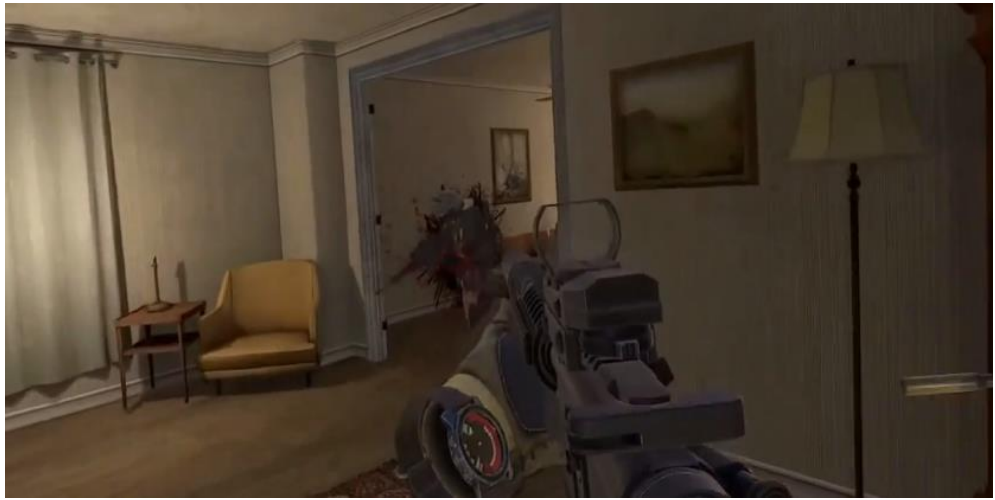

(a)

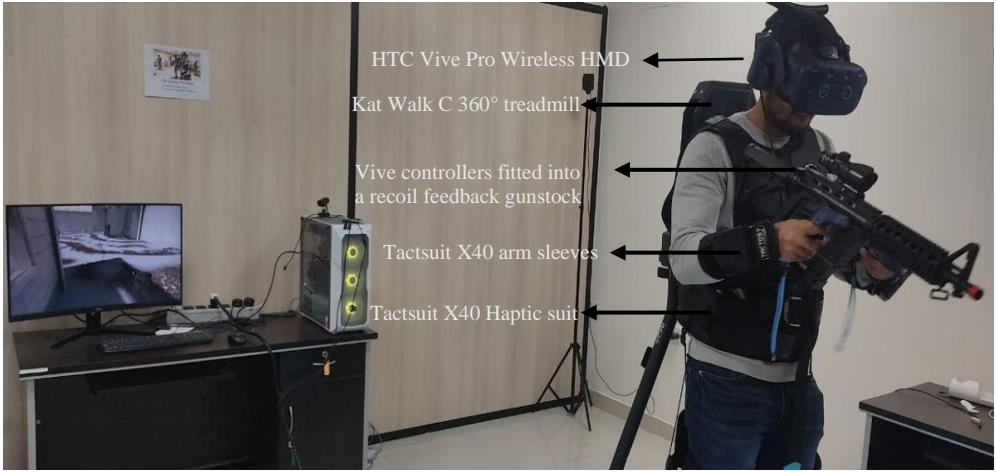

(b)

Figure 1. (a) A desktop view of the interface of the search-and-shoot serious simulation (b) The image of a subject executing the search-and-shoot serious simulation in the HTC Vive Pro HMD VR headset, with the Tactsuit X40 and the Kat Walk C 360º locomotion treadmill integrated (*Disclaimer: Figure 1(b) is for illustrative purposes only and does not represent the actual imagery of the experiment.*)

## 2.3 Experimental design

The participants were equally and randomly divided into two between-subject conditions: experimental condition (haptic feedback, 360° locomotion-integrated VR framework with longitudinal, heterogeneous training) and placebo control (longitudinal, heterogeneous VR training without extrasensory modalities). The entire experiment was 10 days long. After acquiring the participant's consent, demographic details, pre-experiment simulator-sickness-based exclusion questionnaire, sleepiness test, and visual acuity test, the participants were briefed about the objectives to be achieved in the experiment. All the subjects executed the search-and-shoot serious simulation in the 'transfer' complexity level on Day 1 (termed the pre-intervention phase). From Day 2 to Day 9 (termed the intervention phase), the participants alternatively executed the 'rookie', 'intermediate-L1', 'intermediate-L2', and 'professional' variations of the complexity levels on each day. This accounted for the variable training instances incorporated in the 'heterogeneity by practice' framework integrated as a part of the longitudinal intervention process in the between-subject conditions. A detailed description of the different complexity levels designed in the simulation is shown in section 2.4. The participants in the experimental condition executed the pre-intervention, intervention, and post-intervention phases of the search-and-shoot simulation on the HTC Vive Pro VR HMD headset integrated with the Tactsuit X40 haptic feedback suit and the Kat Walk C 360° locomotion setup. The participants in the placebo control condition executed all the phases of the search-and-shoot simulation on the HTC Vive Pro VR HMD headset without the integration of the extrasensory modalities. On Day 10 (post-intervention phase), the participants again executed the search-and-shoot serious simulation in the 'transfer' complexity level.

On Day 1, after acquiring all the pre-experimental details, all participants were made to execute a 25-minute acclimatization session in a dummy simulation in their respective conditions (Only VR or VR with extrasensory modalities). This was done so that the participants could get used to the level of immersion and resolution in the VR simulation and understand the various proprioceptive intricacies associated with the haptic feedback suit and the 360° locomotion treadmill. Various behavioral measures from the search-and-shoot simulation like the percentage number of enemies killed (calculated by dividing the number of enemies killed/9 (number of enemies in the simulation), time taken to complete the simulation, number of bullets/magazines used, and the rate of decrease in health (calculated by subtracting the initial health from the final health and dividing it by the time taken to execute the simulation) were calculated. These behavioral measures were acquired after the participants completed their simulations on Day 1 and Day 10. Following the simulation, participants completed the National Aeronautics Space Administration Task Load Index (NASA-TLX), a widely utilized tool for gauging participants' subjective workload on a 100-point Likert scale, divided into six sub-scales: mental demand, physical demand, temporal demand, frustration level, level of performance satisfaction, and level of effort [38]. In addition, participants completed the computerized versions of the Presence Questionnaire and the Simulator Sickness Questionnaire (SSQ) [39, 40]. The Presence Questionnaire assessed participants' experiences with the simulation on a 7-point Likert scale across questions pertaining to realism, interface quality, ability to act and examine the simulation environment, self-evaluation of performance, and audio-haptic cues. The SSQ, on the other hand, evaluated general discomfort, fatigue, headache, eye strain, vertigo, and other factors on a scale of 0-3, with 0 indicating no effect and 3 signifying severe impact. The total score was calculated across four categories: nausea, oculomotor sickness, disorientation, and overall score. To analyze the impact of different types of between-subject training conditions (heterogeneous training in VR, heterogeneous training in VR integrated with extrasensory modalities) and duration of training (Day 1 and Day 10) on various behavioral, workload, usability, and cybersickness measures, one-way ANOVAs were conducted. Additionally, mixed ANOVAs were performed to evaluate the interaction effects between these variables.

## 2.4 Variation in the simulation complexity

Table 1 shows the variation in the physical characteristics of the enemy avatars and the participant with respect to the complexity level.

Table 1. Variations in the physical characteristics of the simulation with respect to the various complexity levels

| Physical Characteristic | Rookie-level | Intermediate-L1 level | Intermediate-L2 level | Professional level | Transfer level (Only on Day 1 and Day 10) |
|---|---|---|---|---|---|
| Ammunition /magazines available in the M4 Carbine 5.66mm gun | 5 x 30-round detachable magazines | 4 x 30-round detachable magazines | 3 x 30-round detachable magazines | 2 x 30-round detachable magazines | 3 x 30-round detachable magazines |
| Number of enemies in the simulation | 9 (all assault) | 9 (8 assault, 1 stealth) | 9 (7 assault, 2 stealth) | 9 (6 assault, 3 stealth) | 9 (6 assault, 3 stealth) |
| Bullets needed to kill an enemy | Headshot-1<br>Body shot - 5 | Headshot-1<br>Body shot - 5 | Headshot-1<br>Body shot - 5 | Headshot-1<br>Body shot - 5 | Headshot-1<br>Body shot - 5 |
| Health reduction of participant/shot by the enemy | Headshot – 100<br>Body shot - 10 | Headshot – 100<br>Body shot - 20 | Headshot – 100<br>Body shot – 25 | Headshot – 100<br>Body shot - 50 | Headshot – 100<br>Body shot - 20 |
| Proportion of the probabilities of the assault enemies moving towards the participant and the nearest security base | 0.5:0.5 | 0.6:0.4 | 0.7:0.3 | 0.8:0.2 | 0.5:0.5 |
| Distance (in Unreal Engine physical mesh metrics) within which the enemies make the decision to head towards the participant or the security base | 100 Unreal Engine units | 125 Unreal Engine Units | 150 Unreal Engine Units | 175 Unreal Engine Units | 175 Unreal Engine Units |
| Distance (in Unreal Engine physical mesh metrics) within which the enemies make the decision to head towards the participant or the security base | 10 Unreal Engine Units | 30 Unreal Engine Units | 50 Unreal Engine Units | 70 Unreal Engine Units | 50 Unreal Engine Units |

As shown in Table 1, the enemies were divided into two categories: assault and stealth. The assault enemies were programmed to act aggressively by actively moving towards the participant or the nearest security base after spawning at a random location at the beginning of the simulation. The stealth enemies were programmed to remain in hiding in one of the cover-up spots in the simulation and only shoot when the participant was in their vicinity.

## 3. RESULTS

### 2.3 Behavioral measures

2.3.1 Percentage number of enemies killed

The percentage number of enemies killed did not significantly vary across the training conditions ($F(2, 57) = 1.12$, $p = 0.85$, $\eta_p^2 = 0.02$). Across both the training conditions, the percentage number of enemies killed was significantly higher on Day 10 compared to Day 1 (Day 1: $\mu = 56.42$ < Day 10: $\mu = 78.75$; $F(1, 57) = 34.45$, $p < 0.05$, $\eta_p^2 = 0.12$). In addition, the interaction between the type of training conditions and the training phases did not significantly influence the percentage number of enemies killed ($F(2, 57) = 2.49$, $p = 0.79$, $\eta_p^2 = 0.06$).

### 2.3.2 Total time taken to complete the simulation

The total time taken to complete the simulation did not significantly vary across the training conditions ($F (2, 57) = 3.03$, $p = 0.46$, $\eta_p^2 = 0.07$). Across both the training conditions, the total time taken to complete the simulation was significantly lower on Day 10 compared to Day 1 (Day 1: $\mu = 278.34s$ > Day 10: $\mu = 223.45s$; $F (1, 57) = 85.51$, $p < 0.05$, $\eta_p^2 = 0.28$). Besides, the interaction between the type of training conditions and the training phases did not significantly influence the total time taken to complete the simulation ($F (2, 57) = 1.56$, $p = 0.83$, $\eta_p^2 = 0.03$).

### 2.3.2 Rate of decrease in health

The rate of decrease in health did not significantly vary across the training conditions ($F (2, 57) = 3.61$, $p = 0.51$, $\eta_p^2 = 0.1$). Across both the training conditions, the rate of decrease in health was significantly lower on Day 10 compared to Day 1 (Day 1: $\mu = 0.68$ > Day 10: $\mu = 0.34$; $F (1, 57) = 29.56$, $p < 0.05$, $\eta_p^2 = 0.16$). Besides, the interaction between the type of training conditions and the training phases did not significantly influence the rate of decrease in health ($F (2, 57) = 2.33$, $p = 0.61$, $\eta_p^2 = 0.05$).

## 2.3 Measures from the presence questionnaire

The 'realism' subscale in the presence questionnaire was significantly different across different training conditions ($F (2, 57) = 19.54$, $p < 0.05$, $\eta_p^2 = 0.21$). However, across both the training conditions, the realism subscale was not significantly different on Day 10 compared to Day 1 ($F (1, 57) = 4.1$, $p = 0.14$, $\eta_p^2 = 0.09$).

Furthermore, the interaction between the type of training condition and the phase of training significantly influenced the realism subscale ($F (2, 57) = 13.41$, $p < 0.05$, $\eta_p^2 = 0.17$) (See Figure 2). Paired comparisons revealed that there was a significant difference in the realism subscale across both training conditions on Day 1 (VR + Extrasensory modalities: $\mu = 30.56$ > Only VR: $\mu = 22.12$ ($p < 0.05$). Similarly, the realism subscale was proportionately, significantly higher in the VR + Extrasensory modalities training condition compared to the only VR training condition on Day 10 (VR + Extrasensory modalities: $\mu = 32.34$ > Only VR: $\mu = 24.51$ ($p < 0.05$).

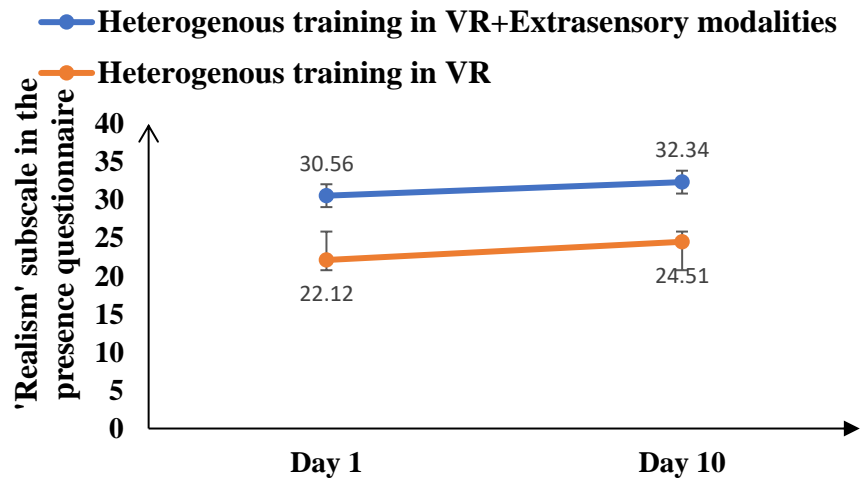

Figure 2. The realism subscale of the presence questionnaire across different training conditions and pre- and post-intervention phases of the experiment (Day 1 and Day 10). The error bars show 95% CI across point estimates.

The 'possibility to examine' subscale in the presence questionnaire was significantly different across different training conditions ($F (2, 57) = 15.78$, $p < 0.05$, $\eta_p^2 = 0.19$). However, across both the training conditions, the 'possibility to examine' subscale was not significantly different on Day 10 compared to Day 1 ($F (1, 57) = 1.12$, $p = 0.98$, $\eta_p^2 = 0.01$).

Furthermore, the interaction between the type of training condition and the phase of training significantly influenced the 'possibility to examine' subscale ($F (2, 57) = 17.81$, $p < 0.05$, $\eta_p^2 = 0.21$) (See Figure 3). Paired comparisons revealed that there was a significant difference in the 'possibility to examine' subscale across both training conditions on Day 1 (VR + Extrasensory modalities: $\mu = 17.45$ > Only VR: $\mu = 11.16$ ($p < 0.05$). Similarly, the 'possibility to examine' subscale was proportionately, significantly higher in the VR + Extrasensory modalities training condition compared to the only VR training condition on Day 10 (VR + Extrasensory modalities: $\mu = 24.12$ > Only VR: $\mu = 14.21$ ($p < 0.05$).

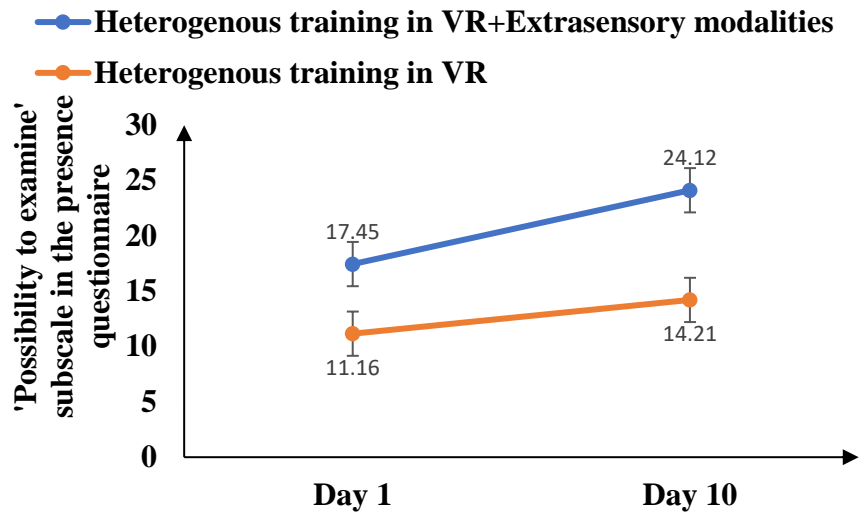

Figure 3. The 'possibility to examine' subscale of the presence questionnaire across different training conditions and pre- and post-intervention phases of the experiment (Day 1 and Day 10). The error bars show 95% CI across point estimates.

All the other variables in the presence questionnaire failed to yield statistical significance.

### 2.4 Measures from the NASA-TLX questionnaire

The 'mental demand' subscale in the NASA-TLX questionnaire was significantly different across different training conditions ($F (2, 57) = 23.45$, $p < 0.05$, $\eta_p^2 = 0.15$). Similarly, across both the training conditions, the mental demand subscale was significantly different on Day 10 compared to Day 1 ($F (1, 57) = 24.33$, $p < 0.05$, $\eta_p^2 = 0.18$).

Furthermore, the interaction between the type of training condition and the phase of training significantly influenced the mental demand subscale ($F (2, 57) = 15.67$, $p < 0.05$, $\eta_p^2 = 0.11$) (See Figure 4). Paired comparisons revealed that there was a significant difference in the mental demand subscale across both training conditions on Day 1 (VR + Extrasensory modalities: $\mu = 85.6$ > Only VR: $\mu = 71.8$ ($p < 0.05$). Similarly, the mental demand subscale was significantly higher in the VR + Extrasensory modalities training condition compared to the only VR training condition on Day 10, even though a drop was recorded in both conditions compared to Day 1 (VR + Extrasensory modalities: $\mu = 78.6$ > Only VR: $\mu = 60.1$ ($p < 0.05$).

Similarly, the 'physical demand' subscale in the NASA-TLX questionnaire was significantly different across different training conditions ($F (2, 57) = 19.62$, $p < 0.05$, $\eta_p^2 = 0.13$). Similarly, across both the training conditions, the physical demand subscale was significantly different on Day 10 compared to Day 1 ($F (1, 57) = 31.05$, $p < 0.05$, $\eta_p^2 = 0.28$).

Furthermore, the interaction between the type of training condition and the phase of training significantly influenced the physical demand subscale ($F (2, 57) = 17.56$, $p < 0.05$, $\eta_p^2 = 0.21$) (See Figure 5). Paired comparisons revealed that there was a significant difference in the physical demand subscale across both training conditions on Day 1 (VR + Extrasensory modalities: $\mu = 81.4$ > Only VR: $\mu = 57.8$ ($p < 0.05$). Similarly, the physical demand subscale was significantly higher in the VR + Extrasensory modalities training condition compared to the only VR training condition on Day 10, even though a drop was recorded in both conditions compared to Day 1 (VR + Extrasensory modalities: $\mu = 69.5$ > Only VR: $\mu = 47.9$ ($p < 0.05$).

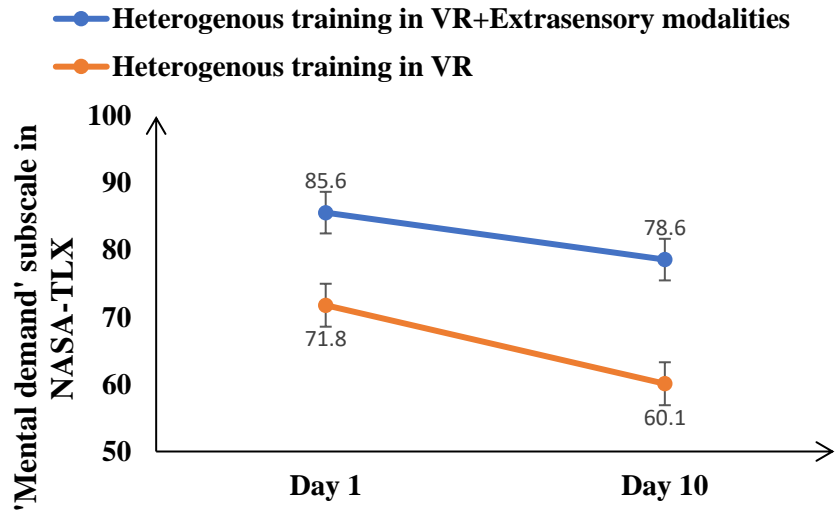

Figure 4. The 'mental demand' subscale of the NASA-TLX questionnaire across different training conditions and pre- and post-intervention phases of the experiment (Day 1 and Day 10). The error bars show 95% CI across point estimates.

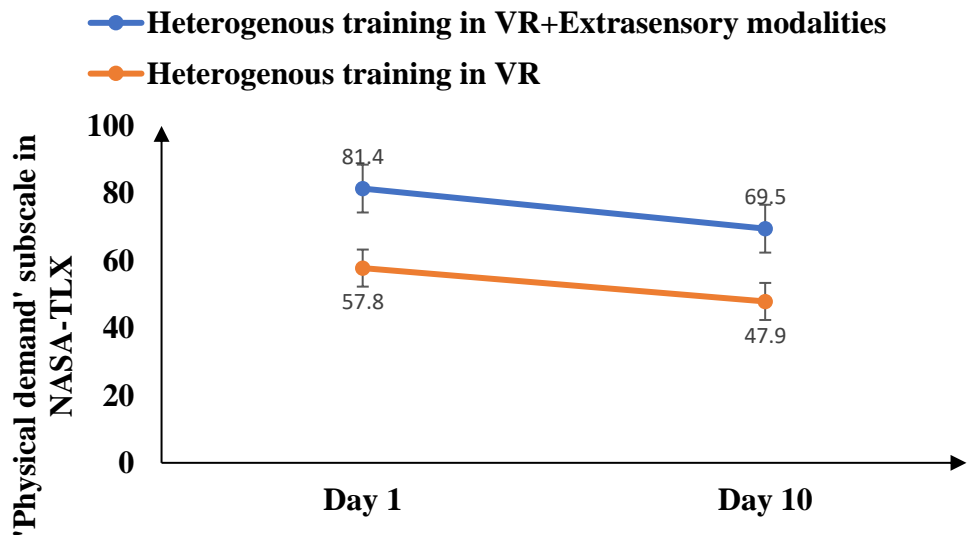

Figure 5. The 'physical demand' subscale of the NASA-TLX questionnaire across different training conditions and pre- and post-intervention phases of the experiment (Day 1 and Day 10). The error bars show 95% CI across point estimates.

The other variables in the NASA-TLX questionnaire (temporal demand, performance satisfaction, level of effort, frustration level) failed to yield statistical significance.

### 2.4 Measures from the SSQ

The 'disorientation' subscale in the SSQ was significantly different across different training conditions ($F(2, 57) = 26.78$, $p < 0.05$, $\eta_p^2 = 0.19$). Similarly, across both the training conditions, the disorientation subscale was significantly different on Day 10 compared to Day 1 ($F(1, 57) = 34.83$, $p < 0.05$, $\eta_p^2 = 0.25$).

Furthermore, the interaction between the type of training condition and the phase of training significantly influenced the disorientation subscale ($F(2, 57) = 13.99$, $p < 0.05$, $\eta_p^2 = 0.17$) (See Figure 6). Paired comparisons revealed that there was a significant difference in the disorientation subscale across both training conditions on Day 1 (VR + Extrasensory modalities: $\mu = 85.4$ > Only VR: $\mu = 71.3$ ($p < 0.05$). Similarly, the disorientation subscale was significantly higher in the

VR + Extrasensory modalities training condition compared to the only VR training condition on Day 10, even though a drop was recorded in both conditions compared to Day 1 (VR + Extrasensory modalities: $\mu = 74.7$ > Only VR: $\mu = 59.4$ ($p < 0.05$).

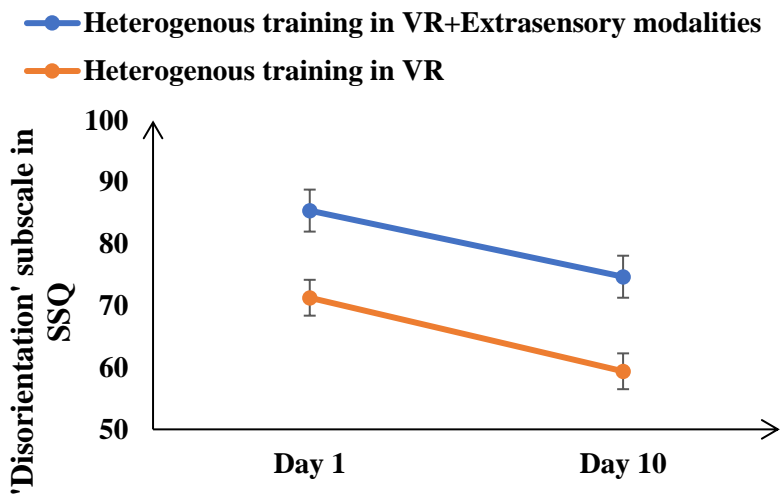

Figure 6. The 'disorientation' subscale of the NASA-TLX questionnaire across different training conditions and pre- and post-intervention phases of the experiment (Day 1 and Day 10). The error bars show 95% CI across point estimates.

The other variables in the SSQ (Nausea subscale, oculomotor sickness subscale, and the total score) failed to yield statistical significance.

## 4. DISCUSSION AND CONCLUSIONS

This study investigated the efficacy of a haptic feedback, 360° locomotion-integrated VR framework with longitudinal, heterogeneous training on decision-making performance in a complex search-and-shoot simulation, compared to a placebo control group undergoing longitudinal, heterogeneous VR training without extrasensory modalities. The experimental design, spanning 10 days with a pre-and post-intervention, aimed to assess the impact of immersive VR training on behavioral, presence, workload, and simulator sickness-based outcomes. Results revealed that there was no significant difference in the behavioral performance parameters between the experimental and the placebo control conditions on Day 10. According to [33], these results could be attributed to the factor of cognitive load and the need for adaptation to the dynamicity of the search-and-shoot simulation, limiting any potential advantage of adding extrasensory modalities to HMD VR-based training. This underlying reason was complemented by the results that were eventually revealed in the experiment: participants in the experimental group experienced enhanced presence and immersion compared to the placebo control group, but they also experienced enhanced mental demand, physical demand, and disorientation. Therefore, even though the participants in the experimental condition experienced enhanced presence, immersion, and the propensity to examine the given serious simulation more freely due to multisensory feedback and the integration of locomotion, it also led to higher workload demands on the part of the participant.

As mentioned above, the results revealed that the participants in the experimental condition experienced enhanced presence and reported an enhancement in their ability to examine the serious simulation more freely, devoid of any spatial restrictions. These results concurred with the work done by researchers in [28], who reasoned that due to the ability of vibrotactile feedback and 360° locomotion's ability to engage the proprioceptive senses, it manages to capture a participant's attention and exacerbates the sense of "being there" in the serious simulation, eventually managing to create a coherent and compelling perceptual experience. Researchers in [26] also reasoned that multisensory feedback was linked to embodied presence, i.e., the feeling of control and influence over the virtual actions being executed by the participants. A qualitative, formal interview with the participants in the experimental condition also revealed that positive emotional responses and excitement also contributed to a sense of immersion and presence, even though it did not translate into higher performance satisfaction.

On the other hand, the results also revealed that the participants in the experimental condition experienced higher mental and physical workload requirements. This was consistent with the work done by researchers in [19], who reasoned that integrating multisensory modalities introduces physical engagement and motor coordination-based challenges in a virtual environment, otherwise absent in a conventional audiovisual setup. With the integration of haptic feedback and 360° locomotion, participants had to navigate through virtual spaces physically, perform physical motor actions (like sprinting and shooting) and respond appropriately to haptic feedback cues. Adding two extra layers of sensory input increased the complexity of information processing, thereby leading to higher mental and physical workload requirements for the participant. These multisensorial cues, according to researchers in [33], also led to sensory conflicts, thereby contradicting the conventional vestibular and proprioceptive feedback, leading to higher disorientation in the experimental condition compared to the placebo control.

Even though this research work is an important contributor towards the understanding of multisensory feedback in VR-based serious simulations and the efficacy of integrating them as a part of the training framework to enhance decision-making performance, it is not without its limitations. While the participants in the experimental condition showed improvements in presence, immersion, and engagement within the VR-based search-and-shoot simulation, the extent to which these skills generalized to the specific decision-making tasks in the search-and-shoot simulation and beyond may have been limited. Furthermore, the experiment design focused on heterogeneous training with varying levels of complexity in the search-and-shoot simulation scenarios. While this approach aimed to challenge participants and enhance their decision-making skills through exposure to diverse environments and enemy behaviors, it may have also led to variability in skill acquisition within the experimental condition. In the future, we plan to incorporate task-specific training and adaptive training tailored to the decision-making demands of the simulation, which might eventually yield more discernible results. In the future, we also intend to improve the sample size and incorporate additional outcome measures (like eye tracking and electroencephalography) to derive neurophysiological interpretations. Future research can build upon these findings by fine-tuning training protocols, emphasizing task-specific training objectives, and exploring additional cognitive psychological factors to enhance decision-making outcomes in VR-based dynamic training environments.